# Fermilab Collider Run II: Accelerator Status and Upgrades

Pushpalatha C. Bhat and William J. Spalding

*Fermi National Accelerator Laboratory*
*Batavia, IL 60510, USA.*

**Abstract.** Fermilab will continue to maintain its pre-eminent position in the world of High Energy Physics, with a unique opportunity to make unprecedented studies of the top quark and major discoveries, until the Large Hadron collider (LHC) at CERN becomes operational near the end of the decade. Run II is well underway with major accelerator and detector upgrades since Run I. A program of further upgrades to the accelerator complex will result in an integrated luminosity of 4-8 $fb^{-1}$ per experiment, by the year 2009.

## INTRODUCTION

Fermilab will remain the High Energy Physics Laboratory at the energy frontier late into the decade, when the Large Hadron Collider (LHC) at CERN begins to operate and the LHC experiments start producing competitive physics results. Therefore, the collider experiments at Fermilab have the unique opportunity for the next several years to make major advances in the study of the top quark that was discovered here in 1995 [1]. The experiments also have a great potential to make major discoveries including new phenomena beyond the Standard Model of particle physics. A rich harvest of standard model precision physics is guaranteed and new insights and surprises are likely, with the unprecedented amounts of data that Run II offers at the highest collision energy ever studied.

The second major proton-antiproton collider run at Fermilab referred to as Run II, started in March 2001. Major upgrades to the accelerator complex include a new 150 GeV proton synchrotron called the Main Injector and a permanent magnet based antiproton storage ring called the Recycler, both housed in a common tunnel. The Main Injector is built as a replacement to the Main Ring (the original machine built in the 6 Km Tevatron tunnel, in 1969) and allows up to 10 times more protons to be injected into the Tevatron as compared to Run I. The Recycler will serve as a second storage ring in the complex, for antiprotons. The collision energy has also been increased from $\sqrt{s}$ =1.80 TeV in Run I to $\sqrt{s}$ =1.96 TeV in Run II. The number of proton and antiproton bunches was increased from six to thirty six for Run II.

A project plan for major additional upgrades, generally referred to as Run II upgrades [2], was developed with the goal of delivering the maximum possible integrated luminosity to the collider experiments, CDF and DØ.

The CDF and DØ detectors also underwent major upgrades in preparation for Run II. In this paper, we do not discuss detector performance and issues, but describe the accelerator performance and the Run II upgrades.

## THE RUN II CAMPAIGN

It is recognized that Run II is a complex campaign of operations, maintenance, upgrades, R&D and studies. This campaign is designed (1) to deliver increasing luminosity in the short-term, by improving the reliability and efficiency of operations and optimizing the operational parameters and (2) to implement and commission a program of upgrades to the accelerator complex to provide significant luminosity increases in the future. The Run II upgrade program consists of two categories of subprojects – Reliability Upgrades and Luminosity Upgrades, and is closely integrated with ongoing operations.

### Reliability Upgrades and Operational Improvements

The strategy for improving reliability is to aggressively pursue the cause of equipment failure either due to individual components or systems. Additionally, upgrade projects address concerns raised in a vulnerability study which identified component failures which, while rare, would result in significant down-times for the complex and loss of integrated luminosity for the experiments. One of the leading causes of lost stores was a defect in voltage to frequency converters (VFC) used in the Tevatron quench protection system. All of the flawed VFC cards were replaced and the quench protection monitoring software was upgraded. The Tevatron RF power amplifiers were desensitized to power glitches. Upgrades to the beam abort logic in the Tevatron and development and implementation of a more sophisticated and robust abort system are being pursued.

Alignment and repairs of components in the accelerators, particularly in the Tevatron, have been undertaken as a major component of the upgrades. During the shutdown in the fall of 2003, a Herculean effort was mounted to re-align the components in the Tevatron. A new alignment network (TevNet) was implemented in the Tevatron to improve alignment precision and to simplify the process of alignment. A large number of dipole cryostats were re-shimmed to correct for cryostat movement, "rolled" magnets were unrolled to reduce corrector magnet currents, deteriorated magnet stands were replaced to improve our ability to align magnets, and real-time motion sensors were installed on Tevatron components to monitor magnet movement remotely. After the shutdown, in a concerted effort to improve operations, dedicated studies were carried out to understand and improve the optics in the Tevatron, improve transfer efficiencies for proton and antiproton bunches for collider stores and for improving the helical orbits. This work resulted in the excellent and continually improving performance of the Tevatron.

## Luminosity Upgrades

The Run II Luminosity Upgrade Program consists of a set of subprojects for upgrades throughout the accelerator complex to increase peak luminosity to about $2.8\times10^{32}$ $cm^{-2}sec^{-1}$. In the following section, we provide a brief discussion of the instantaneous luminosity and the key parameters involved.

### *The Upgrade Strategy*

The integrated luminosity seen by an experiment depends on the peak (instantaneous) luminosity, the luminosity lifetime and reliable operation.

The instantaneous luminosity at each experiment is given by the formula

$$L = \frac{3\gamma f_o B}{\beta^*}\frac{N_p}{\varepsilon_p}\frac{N_{\bar{p}}}{(1+\varepsilon_{\bar{p}}/\varepsilon_p)}H \qquad (1)$$

where $\gamma$ is the relativistic energy factor, $f_o$ is the revolution frequency, $N_p$ and $N_{\bar{p}}$ are the numbers of protons and antiprotons per bunch and $B$ is the number of bunches of each. $\beta^*$ is the beta function at the center of the interaction region, and $\varepsilon_p$ and $\varepsilon_{\bar{p}}$ are the proton and antiproton 95 %, normalized, transverse emittances. $H$ is the hourglass factor due to the bunch lengths.

While the luminosity depends on the transverse emittances explicitly and on the longitudinal emittance through the hourglass factor, the most direct way to increase the luminosity is to increase the proton and antiproton bunch intensities. The term $N_p/\varepsilon_p$ (the proton brightness) is constrained by the maximum tolerable antiproton beam-beam tune shift that it causes. Therefore, the strategy for Run II upgrade is to increase the luminosity primarily by increasing the number of antiprotons and using them efficiently in the collider operations.

The key parameters in the luminosity formula are listed in Table 1, along with parameters defining the rate of antiproton production. The table compares the performance values for present operation with those projected at completion of the upgrade program. It should be noted that the parameters quoted for the Run II Design are for typical average performance.

## The Upgrade Projects

The central strategy for the Run II luminosity upgrade is to increase antiproton production and stack size and to upgrade the Tevatron to handle high intensity bunches. In other words, the complex needs to produce, transport, cool and store more antiprotons and use them efficiently in collider stores in the Tevatron. With stochastic cooling, the stacking rate decreases linearly as the antiproton stack size increases, so a key element of the upgrade plan is to implement electron cooling in a second storage ring, the Recycler, to allow stacking at high rates to very large stack sizes.

**TABLE 1. Key Performance Parameters**

| Parameter | FY04 Best Store | FY04 Last 10 stores Average | Run II *Design* | Units |
|---|---|---|---|---|
| Initial Instantaneous (Peak) Luminosity | 107 | 88 | 284 | $x10^{30}cm^{-2}sec^{-1}$ |
| Integrated Luminosity per week | - | - | 50 | $pb^{-1}$ |
| Store hours per week | - | - | 100 | |
| Store Length | 32 | 27 | 15 | hr |
| Number of Bunches | 36 | 36 | 36 | |
| Number of Protons/bunch | 246 | 249 | 270 | $x10^{9}$ |
| Number of Antiprotons/bunch | 43 | 36 | 131 | $x10^{9}$ |
| β* | 35 | 35 | 35 | cm |
| Effective Transverse Emittance (at collision) | 17 | 17 | 20 | π-mm-mrad |
| Hourglass Form Factor | 0.66 | 0.67 | 0.65 | |
| Antiproton Transmission Efficiency (to low beta) | 86 | 81 | 80 | % |
| Antiproton Stack Used | 181 | 161 | 589 | $x10^{10}$ |
| Avg. Antiproton stacking Rate | 6.8 | 6.4 | 39 | $x10^{10}$ /hr |

The upgrade plan is organized into four main elements:

### *1. Increase number of protons on the antiproton production target*

The primary goal is to double the antiproton production rate by doubling the proton intensity on the production target and to make necessary upgrades to the target station to take full advantage of the increased incident proton flux. The doubling of proton intensity for antiproton production is accomplished by a technique called “slip stacking” [3], which merges two batches of protons from the Booster. The *design* goal is to achieve sustained proton intensities of $8x10^{12}$ protons with specific requirements that bunch length be less than 1.5 nsec and transverse emittance less than 25 π mm-mrad. This requires upgrades to the Main Injector RF system to compensate for beam-loading effects. Other upgrades in the Main Injector that are aimed at improving the Main Injector performance are also planned.

Slip stacking in the Main Injector has been tested for total proton intensities of the order of $7x10^{12}$ protons per pulse and with the specified bunch length on the antiproton production target. It was used for antiproton production and stacking for several days prior to the fall 2004 shutdown and a stacking rate increase of ~15% was observed. This was achieved with only 12 of the 18 RF stations upgraded for beam-loading compensation. The remaining 6 stations will be upgraded during the shutdown and we expect to commission slip stacking at full intensity following the shutdown. Figure 1 shows a mountain range picture and a fast time plot of slip stacking of two booster batches of protons, demonstrating clean merging of the two batches.

To reduce damage to the antiproton target station due to increased flux delivered by slip stacking, new heat-resilient target materials (Stainless Steel targets with graphite cover) have been developed and installed. Additional protection can be provided if necessary by increasing the beam spot size by about 10%. A system for sweeping the beam spot across the target (and a system to compensate for it downstream) has been developed and retained as a back-up.

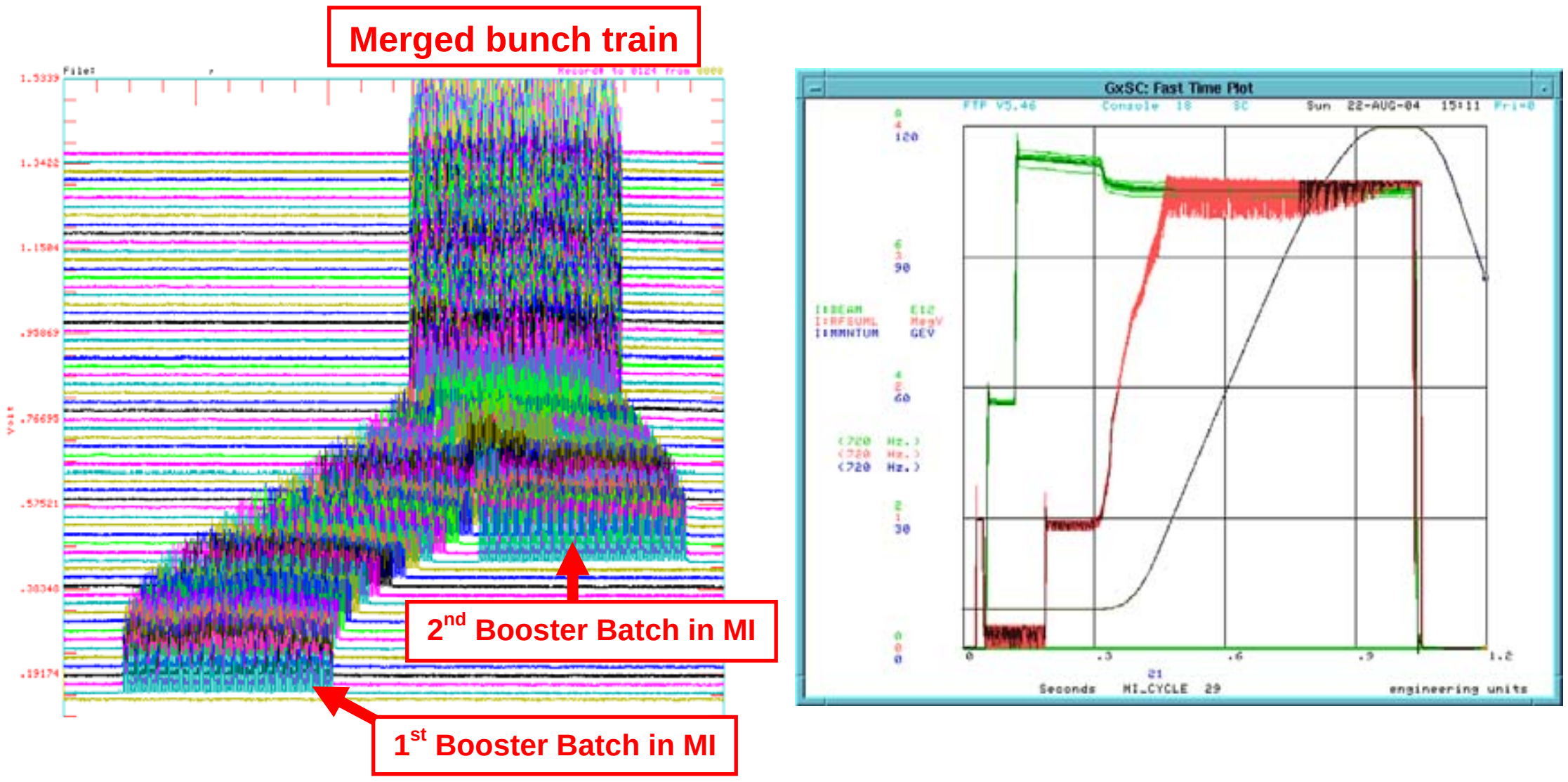

**FIGURE 1.** (Left) Mountain range picture showing merging of two Booster batches of protons in the Main Injector. (Right) Fast time plot during slip stacking – the green trace is the beam intensity (scale 0 to $8x10^{12}$ protons), the red trace is the 53 MHz RF voltage and the black curve shows the beam momentum for acceleration from 8 to 120 GeV.

## *2. Increase Antiproton Collection*

This project is aimed at increasing the acceptance of antiprotons and focuses on the elements that are immediately downstream of the production target - the lithium collection lens, the AP2 beamline and the Debuncher ring. The lithium lens is the first component downstream of the production target and has significant impact on the acceptance of the antiproton source. The number of antiprotons collected at the upstream end of the AP2 beamline increases with the magnetic field gradient in the lens. A new lens design under development will increase operational reliability and longevity and allow operation at a magnetic gradient of 1000 T/m instead of the present 750 T/m, giving 10-15% increase in acceptance. A prototype lens (with improved thermal and mechanical properties and diffusion-bonded Titanium body) has been successfully bench-tested at the higher design gradient.

The AP2 beamline transports the antiprotons from the lithium lens to the Debuncher ring where the antiproton beam is de-bunched and the momentum spread reduced before the antiprotons are stacked and further cooled in the Accumulator. The plan to increase antiproton acceptance in both the AP2 beamline and the Debuncher includes identifying limiting apertures and eliminating them through realignment, relocation, improvements in beam steering, or by rebuilding components.

Beam-based procedures have been developed to identify limiting apertures in the AP2 beamline and the Debuncher. A 26% increase in the combined horizontal aperture ($23\pi$ to $29\pi$) has been seen recently by using bumps in the Debuncher extraction region. Motorized stands for quadrupoles in the Debuncher are being installed for improved local orbit control. Planned re-location of some components and replacement or re-building of some with larger physical apertures, beam-based alignment and orbit tuning, are expected to increase the combined aperture of AP2 and the Debuncher by nearly a factor of two.

### *3. Increase Antiproton Stacking & Storing Capabilities*

This branch of the upgrade project encompasses stacking and cooling of antiprotons both in the antiproton source and the Recycler. The flux capability of the Accumulator stacktail stochastic cooling system will be upgraded to achieve three times the current stacking rate at small stack sizes. To sustain such high rates for several hours and build up large stacks of $600\text{x}10^{10}$ antiprotons, partial stacks will be transferred every 30 minutes from the Accumulator to the Recycler. The antiprotons are stacked and stored in the Recycler until used in a collider shot. The Recycler, which will serve as the second antiproton storage ring, will use both stochastic cooling and electron cooling systems to maintain large stacks with small transverse and longitudinal emittances.

Commissioning the Recycler, ready for installation and implementation of electron cooling, was one of the major milestones for Run II upgrades. This milestone was met, on schedule, on June 1, 2004. This required the Recycler performance to meet several stringent criteria for vacuum, injections and extractions, stochastic cooling and storage of antiprotons. The vacuum performance and the emittance growth due to residual gas Coulomb scattering was identified in 2003 as the most significant obstacle in bringing the Recycler into operations. These issues were successfully resolved during the 2003 shutdown by baking the entire Recycler and adding additional vacuum pumps at strategic locations.

The excellent performance of the Recycler itself and the transfers to and from the machine, has allowed the development of a mixed source mode for collider operation, where antiprotons are injected into the Tevatron from both the Accumulator and the Recycler. About $2650\text{x}10^{10}$ antiprotons have been transferred so far from the Accumulator to the Recycler in a total of 65 transfers. The average transfer efficiency from the Accumulator to Main Injector to Recycler has been above 90%. A novel, new scheme has been developed [4] to selectively isolate and extract the cold beam (close to the synchronous momentum) from the high momentum particles in the beam, for collider operation in the mixed source mode. Mixed source operation produced record luminosity stores prior to the 2004 shutdown.

The next major steps for increased antiproton stacks are (1) implementation and use of electron cooling of antiprotons in the Recycler [5], (2) upgrading the stacktail cooling system in the Accumulator and (3) implementing “Rapid transfers” of antiprotons between the Accumulator and Recycler (1-2 minutes per transfer).

Electron cooling of the 8 GeV antiprotons in the Recycler requires a high quality electron beam (small beam size, small angular spread) with a current of about 500 mA and kinetic energy of 4.3 MeV. A 4.3 MV Pelletron is used to produce this electron beam and the cooling of antiprotons by these electrons takes place in a 20 m section of the Recycler with ten solenoid modules and magnetic field correctors. An R&D program has been completed at the Wide Band Lab at Fermilab to produce an electron beam with the required properties, and the equipment is currently being installed in the Recycler. The electron cooling of antiprotons is expected to be demonstrated by mid-2005 and to be contributing to collider operation by the end of 2005.

Once electron cooling of antiprotons is operational, the accumulator stacktail cooling system will be upgraded in two stages to provide higher stacking rates. The changes to the stacktail system require the large stacks no longer be maintained in the Accumulator, and so depend on the Recycler with its electron cooling. In the first stage, the existing cooling tanks will be reconfigured to increase the stacking rate for small stacks to ~$30\times10^{10}$ per hour. The stack will be transferred to the Recycler every 1-2 hours and stored and cooled with electron cooling. This upgrade is quick and reversible if problems arise operating electron cooling. After significant operational experience is gained with the higher stacking rates, with electron cooling in the Recycler, and with rapid transfers, the band-width of the stacktail cooling system will be upgraded from 2-4 GHz to 2-6 GHz. This upgrade requires new pickups and kickers in the Accumulator, and is therefore not easily reversible. It is expected to allow zero stack stacking rates in excess of $45\times10^{10}$ per hour, and is planned for the 2006 shutdown.

With the bandwidth upgrade it will be necessary to transfer a partial-stack of antiprotons from the Accumulator to the Recycler every half hour. A transfer time of 1-2 minutes (during which stacking in the antiproton source is interrupted) is to be achieved by automating the transfer process, which will require upgrades to the beam line instrumentation, improvements to the power supply regulation, and feedback from the Main Injector damper system. These upgrades are currently in progress and are expected to be completed by mid-2005.

### *4. Upgrade the Tevatron to handle higher intensity bunches*

This fourth branch of the upgrades will optimize the performance of the Tevatron with the increased bunch intensities provided by the upgrades discussed above. The upgrades to the Tevatron include improvements to the proton and antiproton orbits, major improvements to a variety of instrumentation systems and mitigation of undesirable beam-beam effects through passive and active compensation methods.

A number of instrumentation upgrades will improve beam diagnostics in the Tevatron. They include fabrication and implementation of a 1.7 GHz Schottky detector, tune tracker, SyncLite monitor (using synchrotron light to measure beam profiles), abort gap monitor and head-tail monitor. A new Beam Position Monitor system will provide (1) an order of magnitude improvement in position resolution for closed orbit measurements and (2) simultaneous measurement of proton and antiproton beam positions. This upgrade is expected to be completed in early 2005. An Ionization Profile Monitor will help measure injection mismatch and emittance evolution on the ramp. Another major instrumentation upgrade is that of the beam loss monitors for the Tevatron abort system.

Proton and antiproton orbits in the Tevatron are separated helices to avoid collisions at locations other than CDF and DØ. However, long-range interactions between the proton and antiproton bunches can lead to significant tune shifts and shorten luminosity lifetime. In general, these tune shifts decrease as the second power of the orbit separation, so an increase in this separation can significantly reduce such detrimental beam-beam effects. The existing electrostatic separators are now operated at higher voltages for increased separation. Additional separators are being installed, and polarity switches are being added to allow more flexibility in the helix operation.

An electron lens [6] was installed in the Tevatron in 2001 with the intent of testing its use for active beam-beam compensation. However, it has been primarily used, so far, as an abort gap cleaner. Preliminary beam studies carried out last year yielded promising results for bunch-by-bunch tune corrections. A second upgraded electron lens will be installed in 2005.

## PRESENT PERFORMANCE OF THE COMPLEX

The excellent collider performance during the fiscal year 2004 (FY04) is a direct result of the reliability and operational improvements and upgrades - the alignment work in the Tevatron, optics and helix improvements, measures implemented throughout the complex for improved reliability, and concerted efforts on improving operating efficiency. Because of improved reliability of the Tevatron, it was possible to run longer stores and provide time to accumulate large antiproton stacks, which in turn produced high peak (initial instantaneous) luminosities. Improved transfer efficiency of the beams from the Main Injector to the Tevatron and shorter bunches (low longitudinal emittance) due to the use of dampers [7] were also important factors in the improved luminosity performance. The peak luminosity record was ~$1.1 \times 10^{32}$ $cm^{-2}sec^{-1}$, exceeding the design specification for the operation of the complex with the Main Injector. The four highest peak luminosity stores resulted from mixed source operation. The record weekly integrated luminosity delivered to the experiments in FY04 was $18.7 pb^{-1}$. The peak luminosities for stores during FY02, FY03 and FY04 are shown in Figure 2. The weekly and the total integrated luminosities delivered so far during Run II are shown in Figure 3. Figure 4 shows integrated luminosity during FY02, FY03 and FY04 and projections for FY04. The delivered integrated luminosity exceeded the design goal and doubled since FY03.

While other performance parameters have exceeded the design goals for FY04, the antiproton stacking rate was typically ¾$^{th}$ of the goal set for the end of the year. This reduced performance of the stacking system is presently under investigation. Initial measurements suggest alignment or optics matching problems between the Debuncher and Accumulator, and this is a focus of attention for work in the present shutdown, which started at the end of August 2004.

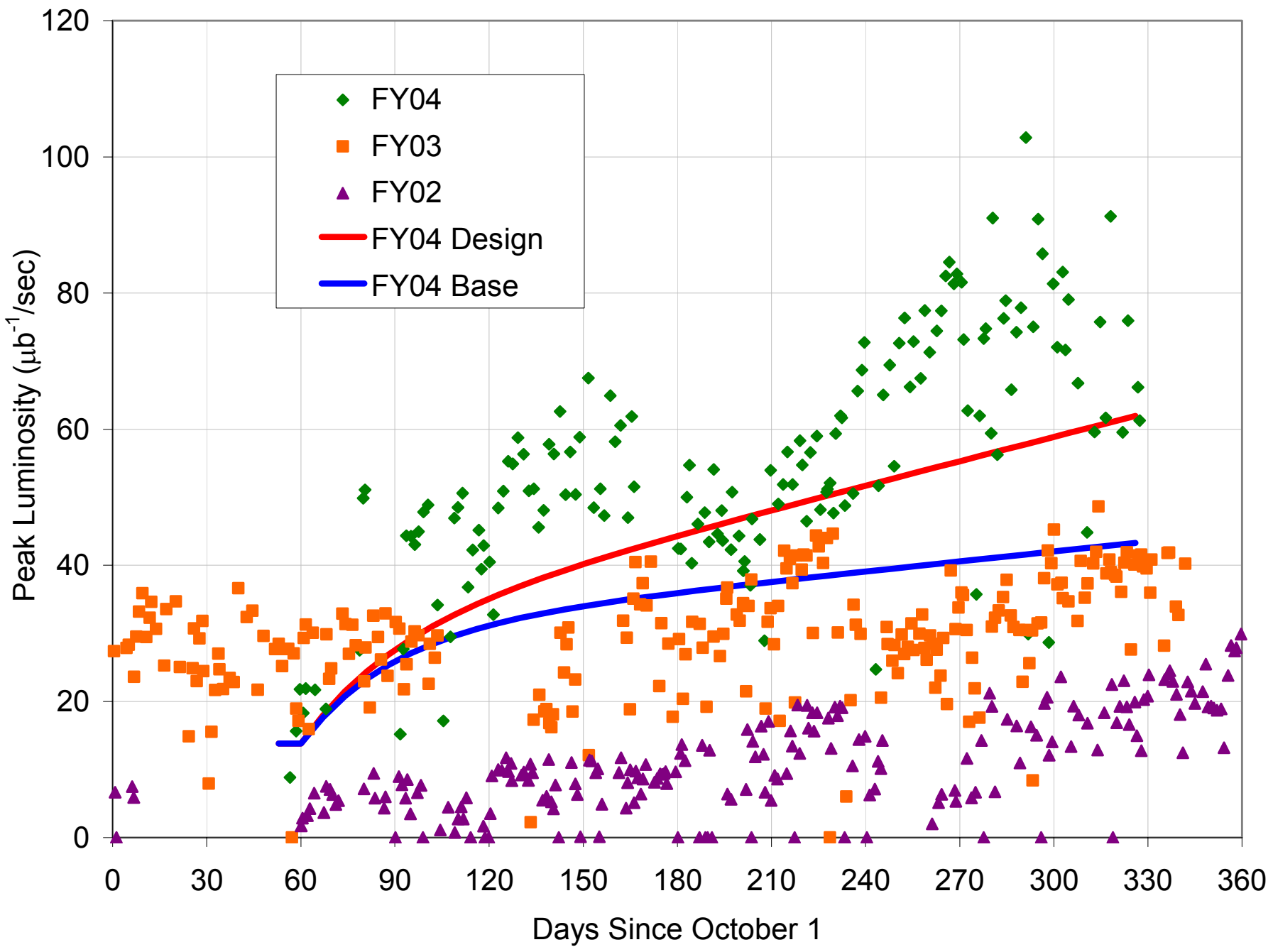

**FIGURE 2.** The peak luminosities obtained for collider stores in FY02, FY03 and FY04, and comparisons with FY04 design and base goals (see discussion on luminosity projections).

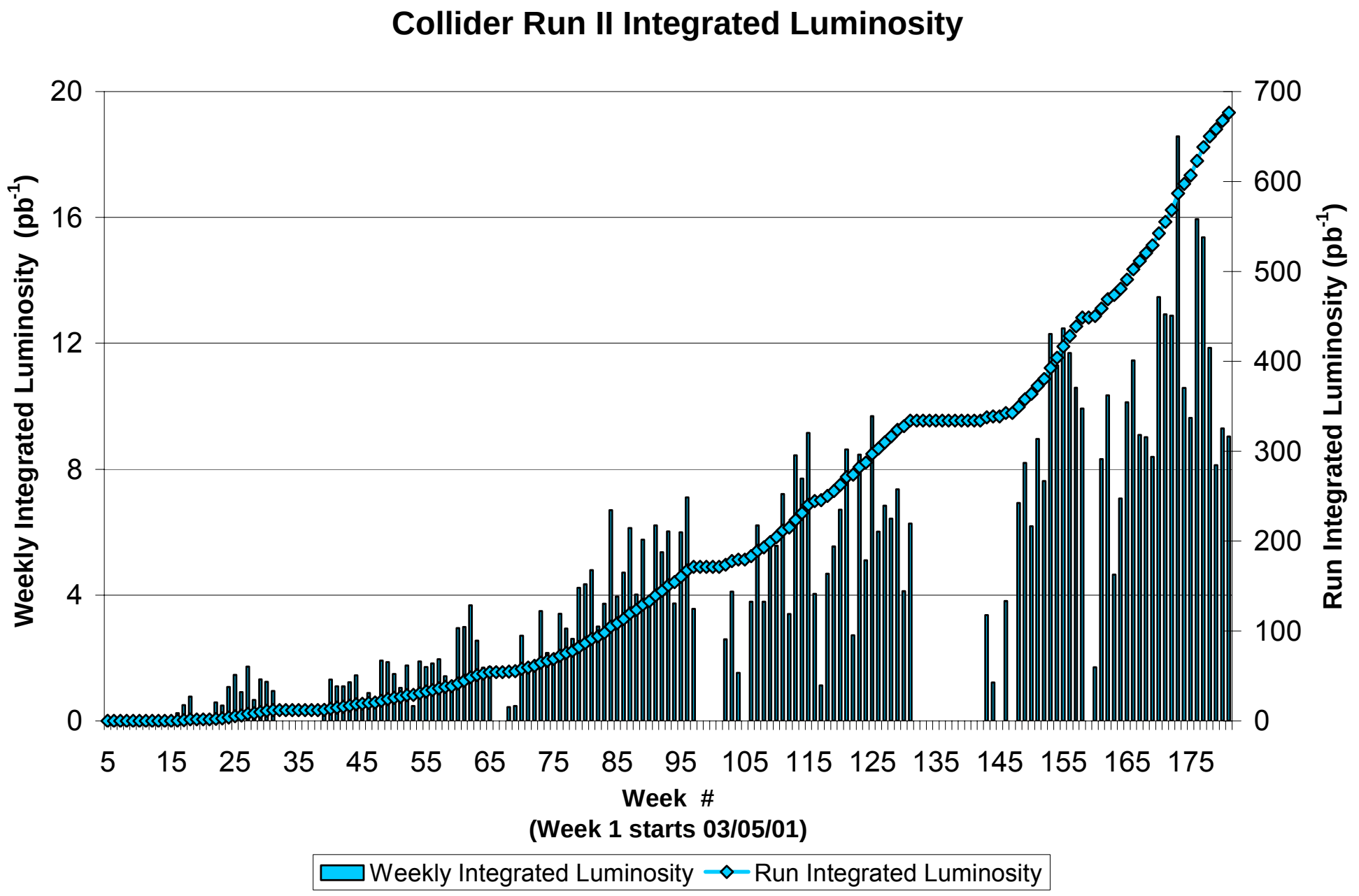

**FIGURE 3.** The weekly and total integrated luminosities delivered to the collider experiments in Run II, through FY04.

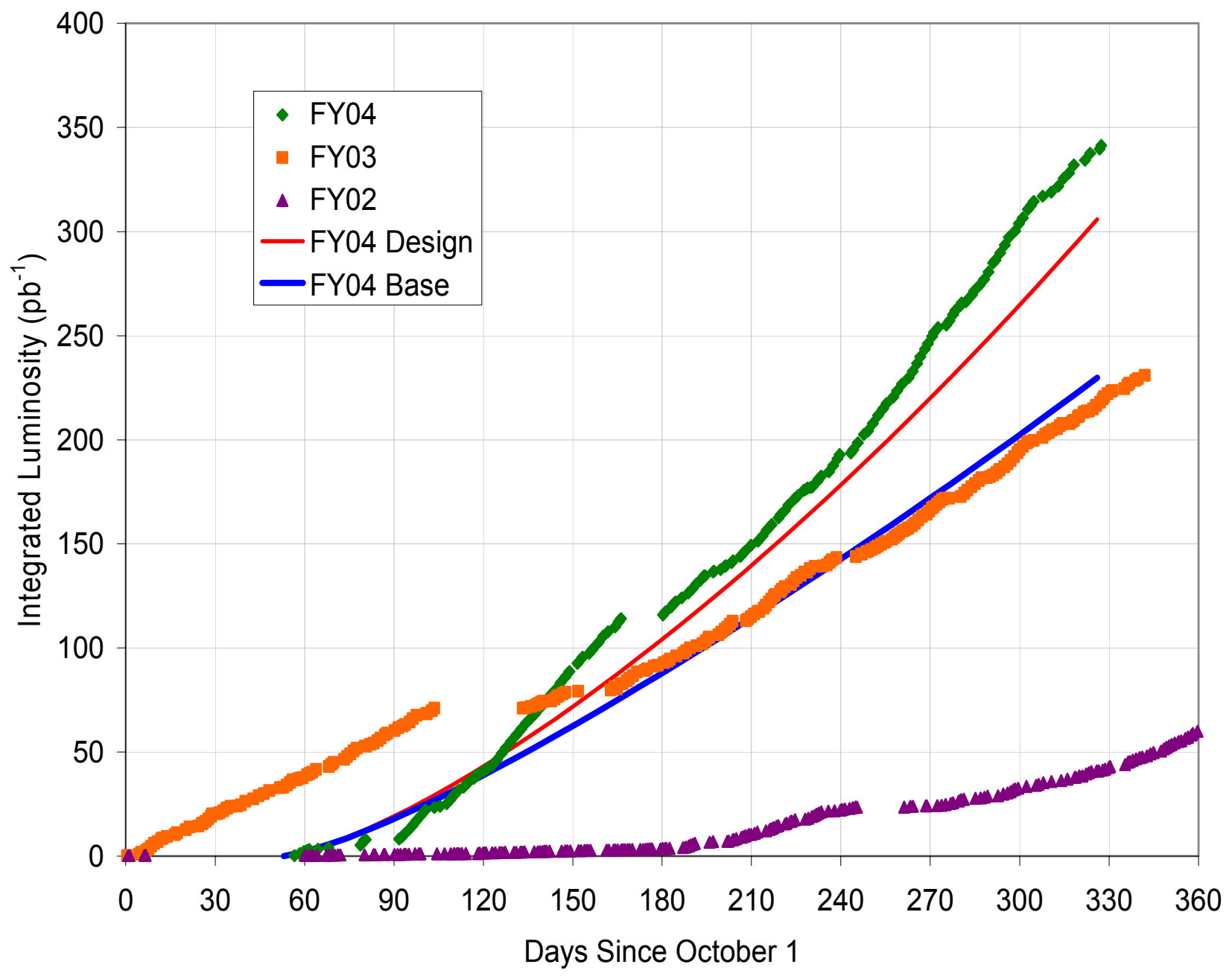

**FIGURE 4.** Comparison of delivered total integrated luminosity for FY02, FY03 and FY04, with design and base projections for FY04.

# LUMINOSITY PROJECTIONS

A parametric model is used to make luminosity projections. The model incorporates three classes of parameters.

(1) Performance parameters from simulation and from operational data. These include antiproton stacking rate, transfer efficiencies and bunch intensities.

(2) Operating scenario parameters from current operating experience. These define operating efficiency and estimated downtime due to quenches and equipment failure.

(3) Learning rates when new upgrades are introduced, and recovery rates after each scheduled shutdown.

Two luminosity projections are presented here. The *design projection*, represents the expected average performance of the complex with the successful completion of the upgrades as planned. Engineering design margin is maintained for each sub-project. The *base projection* illustrates a conservative fall-back scenario in which electron cooling is not successful and therefore the stacktail upgrade cannot be implemented. Additionally it includes only minimal gain in the Debuncher to Accumulator transfers and no gain from mixed source operation. These projections are illustrated in Figures 5 and 6 in terms of weekly integrated luminosity and total integrated luminosity through 2009.

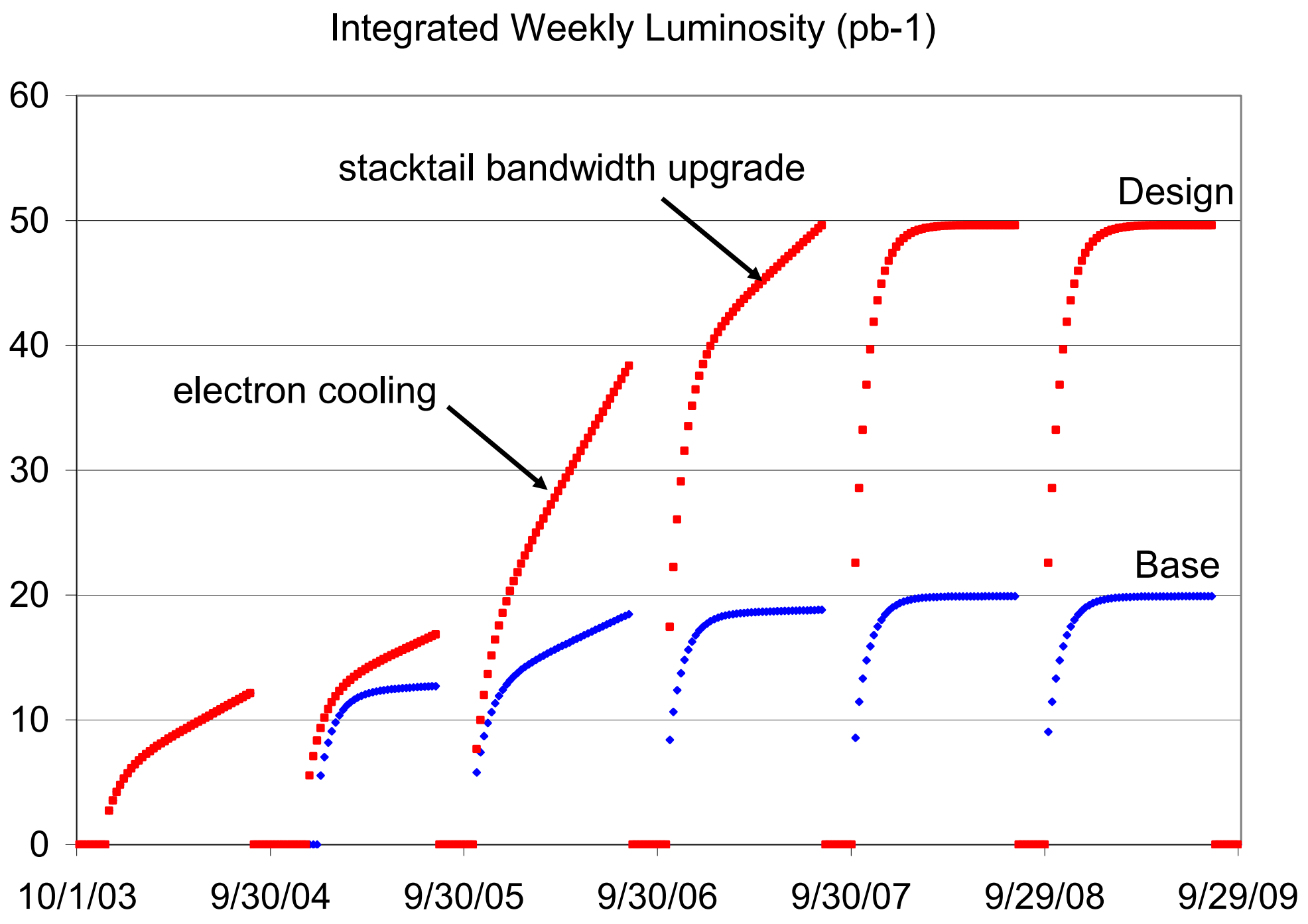

**FIGURE 5.** Integrated weekly luminosity projections through FY09. The red points show the design projection and the blue points the base projection (see text for details).

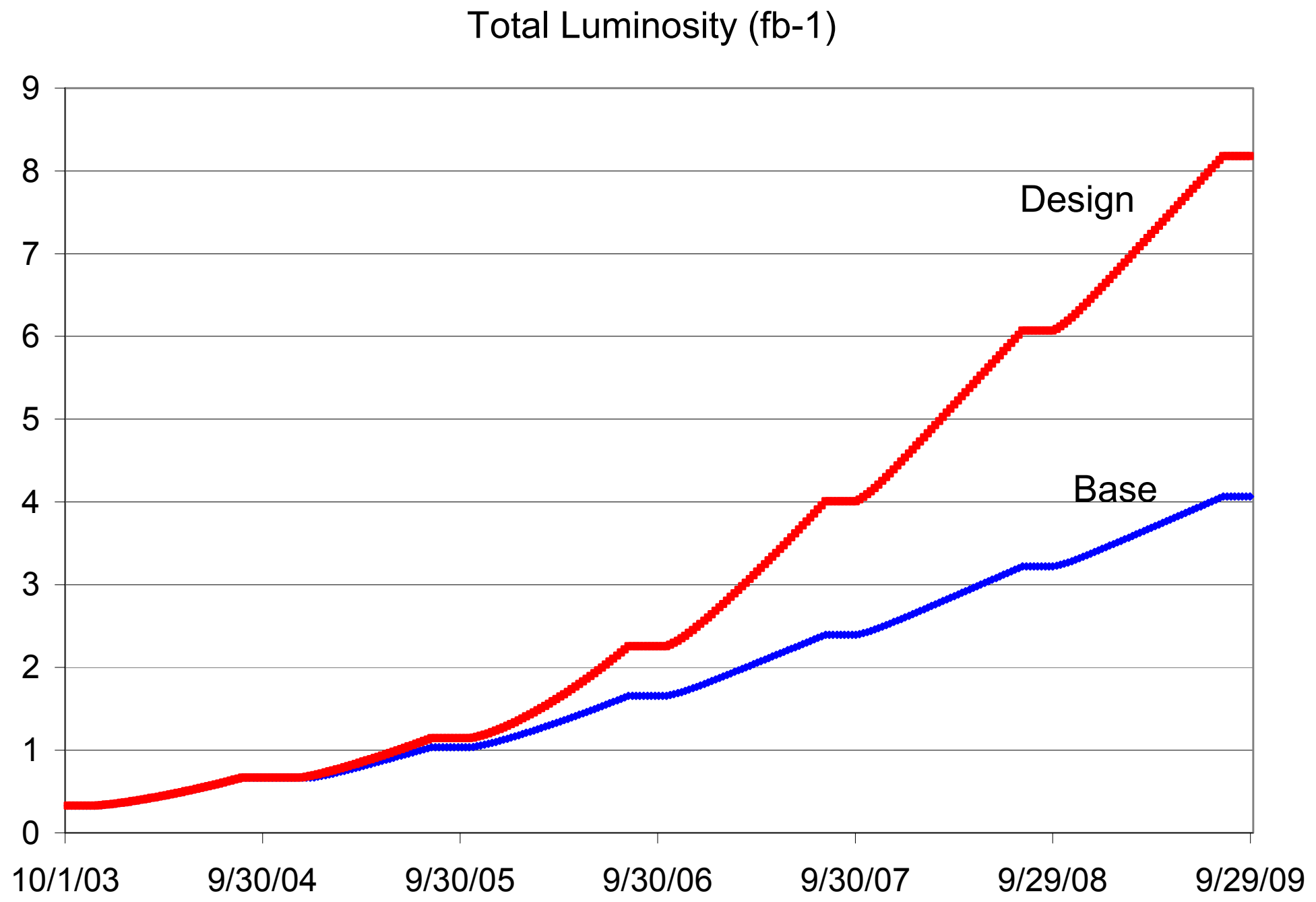

**FIGURE 6.** Total integrated luminosity projections through FY09, for the scenarios discussed.

## SUMMARY

Fermilab Run II is progressing well with 670 $pb^{-1}$ delivered to the experiments so far, and the accelerator has outperformed the design projection for FY04. The peak and weekly luminosities have doubled since a year ago. A complex-wide plan that integrates operations and reliability improvements and luminosity upgrade projects is in place. Technical progress on the upgrades has been excellent. The recently updated plan projects an integrated luminosity in excess of 8 $fb^{-1}$ per collider experiment by the end of FY09, and fall-back scenarios predict integrated luminosities above 4 $fb^{-1}$. These projections assure a rich harvest of precision physics results from CDF and DØ experiments and an excellent opportunity for major discoveries, if nature has new physics within the reach of the Tevatron energy.

## ACKNOWLEDGMENTS

Many people are working extremely hard on all aspects of Run II. In particular we would like to acknowledge those working on upgrades, including support from all divisions at Fermilab and help from the experiments and other institutions. We thank David McGinnis for enlightening discussions and many contributions, and the project leaders for their comments. Fermilab is operated by the Universities Research Association under contract number DE-AC02-76CH03000 with the U.S. Department of Energy.

## REFERENCES

1. CDF Collaboration, F. Abe *et al.*, Phys. Rev. Lett., 74 (1995) 2626; DØ Collaboration, S. Abachi, *et al.*, Phys. Rev. Lett., 74 (1995) 2632; For a comprehensive review of top quark physics at the Tevatron, see P.C. Bhat, H.B. Prosper and S.S. Snyder, Int. J. Mod. Phys. A13 (1998) 5113.
2. The Run II Luminosity Upgrade at the Fermilab Tevatron: Project Plan and Resource-Loaded Schedule, http://www-bdnew.fnal.gov/doereview03/RunII_Upgrade_Plan_v1.0.pdf; The Run II Luminosity Upgrade at the Fermilab Tevatron, v2.0 Project Plan and Resource-Loaded Schedule, http://www-bdnew.fnal.gov/doereview04/RunII_Upgrade_Plan_v2.0.pdf
3. K.Koba, *et al.*, Proceedings of the IEEE Particle Accelerator Conference 2003, p1736, and references therein.
4. C. M. Bhat, *Physics Letters A*, Vol. 330 (2004) 481.
5. J. Leibfritz, *et al.*, to be published in the Proceedings of the European Particle Accelerator Conference 2004.
6. V. Shiltsev, to be published in the Proceedings of the European Particle Accelerator Conference 2004.
7. G.W. Foster, *et al.*, Proceedings of the IEEE Particle Accelerator Conference 2003, p323.